\documentclass{camera} 
\usepackage{graphicx}  
\usepackage{amssymb}
\begin{document}
%
%
\title{JET PHYSICS AT TEVATRON}

%
\author{Giuseppe Latino\footnote{latino@fnal.gov}\\
({\em On behalf of the CDF and $D{\not}O$ Collaborations})}
%
%
\organization{Department of Physics and Astronomy, University of New Mexico, \\ 
800 Yale Blvd. NE, Albuquerque NM 87131, USA }

\maketitle

%
\begin{center}
{\bf Abstract}
\end{center}
An overview of Run I jet physics at the $p\bar p$ Fermilab Tevatron Collider 
with a particular emphasis on inclusive jet cross section measurements is given. 
The impact of these studies on PDFs constrain from global fits is underlined. 
Preliminary results on inclusive jet and di-jet mass cross section measurements 
in Run II are then summarized. 

\vspace{-0.4cm}
\section{Introduction}
\vspace{-0.2cm}

Jet measurements at the Fermilab Tevatron Collider represent a 
fundamental test of next-to-leading order perturbative QCD 
(NLO pQCD) 
at the highest center-of-mass energy so far probed. 
Consequently, they are potentially sensitive to new phenomena beyond the 
Standard Model, the existence of new particles as well as of new 
interactions down to a distance scale of $\sim10^{-19}$ m being expected to be 
revealed by deviations from theory. 
Furthermore, these measurements allow the extraction on a wider kinematical range 
of the fundamental parameters of QCD: ${\alpha}_s$ and the parton distribution 
functions (PDFs). 
\vspace{-0.4cm}
\section{Run I Results: Inclusive Jet Cross Section}
\vspace{-0.2cm}
During Run I, the CDF~\cite{CDF_I} and $D{\not}O$~\cite{D0_I} Collaborations performed
several jet measurements as part of their physics program at the Tevatron.
Stringent tests of LO and NLO pQCD were obtained from the study of final states with
high $E_T$ jets: inclusive jet and di-jet differential cross sections, di-jet mass, di-jet 
angular and multijet distributions as well as the Z/W + jets production cross sections were 
among the most significant measurements to be compared with theory. 
In general, QCD predictions gave a reasonable description of the observed data. 
\begin{center}
\begin{figure}[!htb]
\vspace{-1.cm}
\includegraphics[width=0.5\linewidth,angle=0]{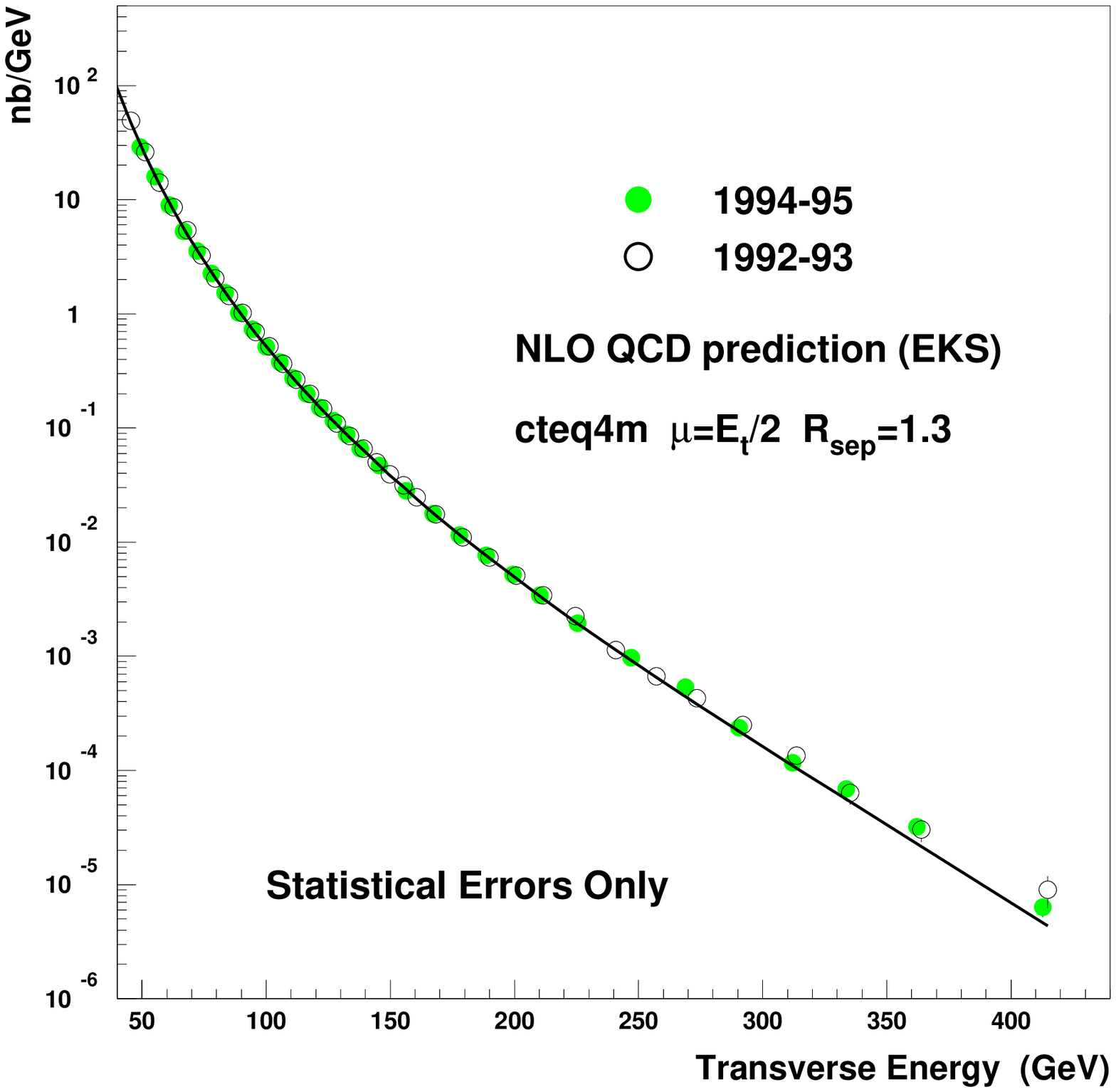}
\includegraphics[width=0.5\linewidth,angle=0]{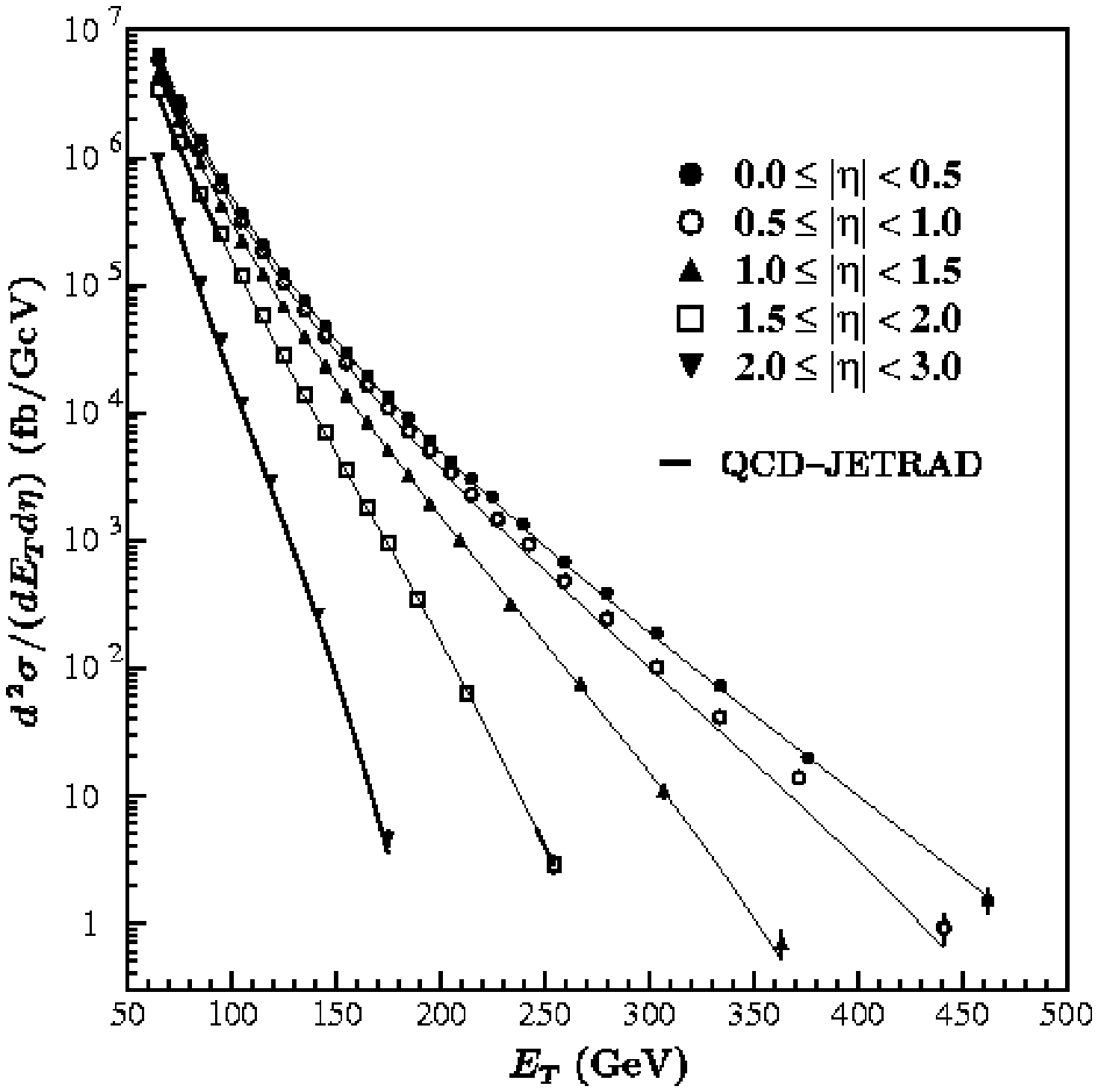}
\vspace{-0.9cm}
\caption{
{\em Inclusive jet cross section as measured in CDF~\cite{CDF_new_incl_jet} (left)
and $D{\not}O$~\cite{D0_new_incl_jet} (right) respectively using 87 and 95
$pb^{-1}$ of Run I data at $\sqrt{s}$ = 1.8 TeV.}
}
\label{fig:fig01}
\vspace{-0.8cm}
\end{figure}
\end{center}

However, great interest was raised when initial inclusive jet cross section
measurements showed an excess of data at high $E_T$ over NLO theoretical predictions 
using then-current PDFs~\cite{CDF_old_incl_jet}. Global PDF analyses by CTEQ~\cite{CTEQ_PDF} 
have demonstrated that such excess can be explained in terms of a larger than expected gluon 
distribution at high $x$, a kinematical region were it is not yet well constrained.
Subsequent measurements performed with an increased data 
sample~\cite{CDF_new_incl_jet,D0_new_incl_jet} (see fig.~\ref{fig:fig01})
showed a better data-theory agreement when PDFs with an increased high $x$ 
gluon contribution (CTEQ4HJ) were considered in the NLO calculations. Given the flexibility 
allowed by current knowledge of PDFs, CDF and $D{\not}O$ data were found to be consistent 
between them, with previous results and with NLO pQCD. 

\begin{center}
\begin{figure}[!htb]
\vspace{-0.5cm}
\includegraphics[width=0.5\linewidth,angle=0]{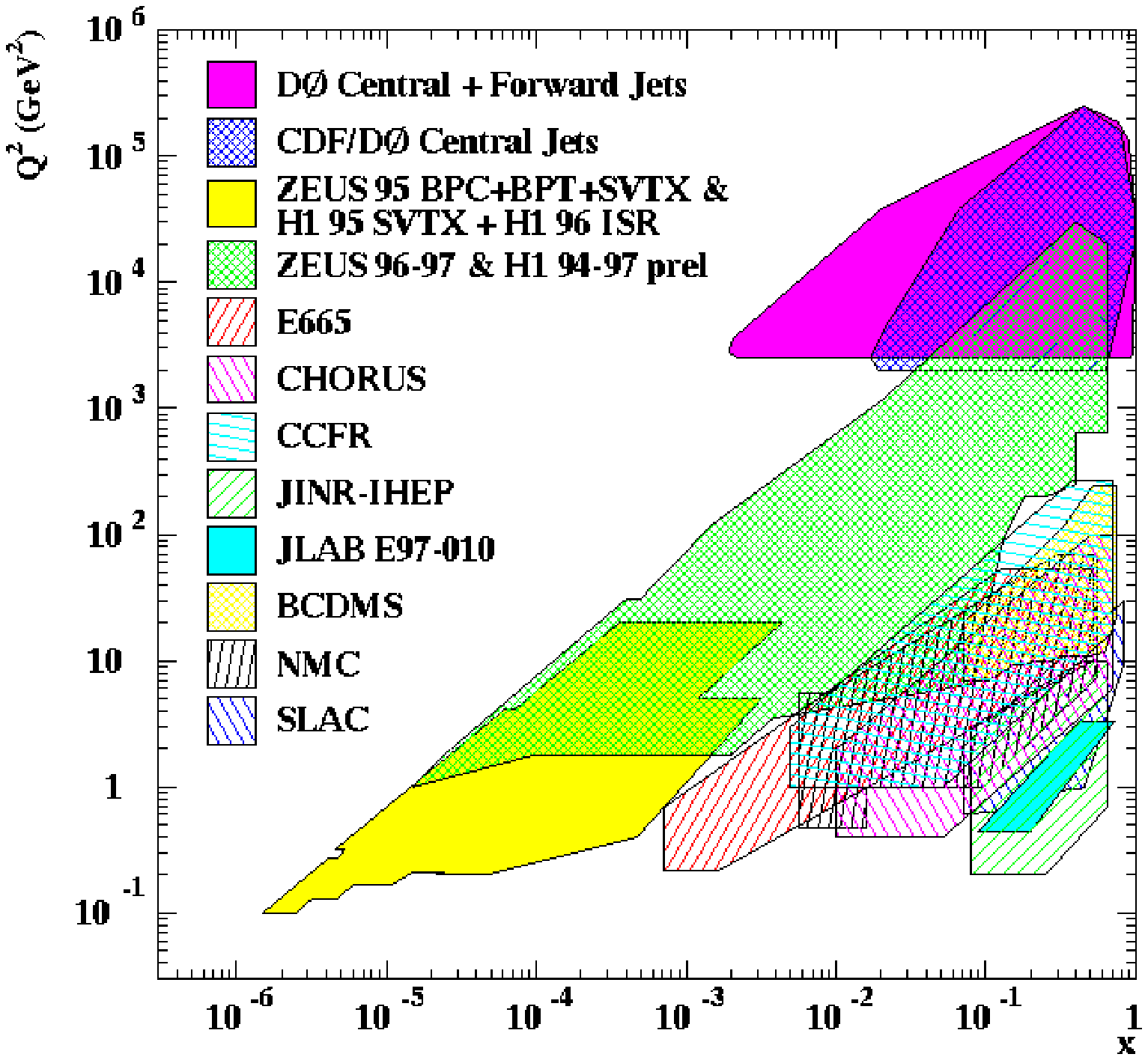}
\includegraphics[width=0.5\linewidth,angle=0]{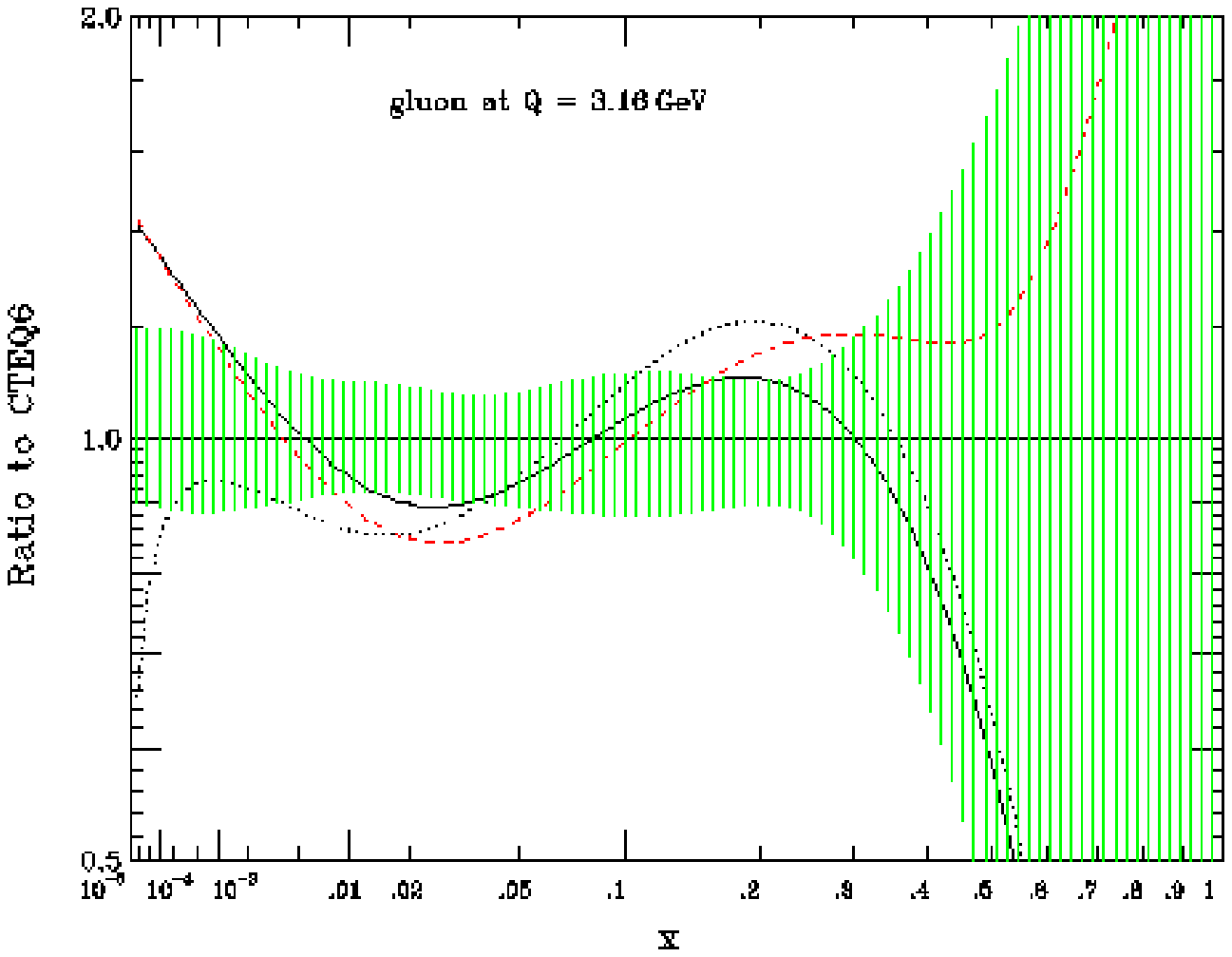}
\vspace{-0.9cm}
\caption{
{\em Left: kinematical reach of CDF and $D{\not}O$ in the $x-Q^2$ plane as compared to other collider 
and fixed target experiments~\cite{D0_new_incl_jet}. Right: uncertainty band for the gluon 
distribution function at $Q^2$ = 10 GeV$^2$ in 
the CTEQ6M set~\cite{CTEQ6}; comparisons to CTEQ5M1 (solid), CTEQ5HJ (dashed) and MRST01 
(dotted) are also shown.} 
}
\label{fig:fig02}
\vspace{-1.0cm}
\end{figure}
\end{center}

Furthermore, results from these analyses have been used in most recent global PDFs fits where 
an enhanced statistical weight was given to Tevatron high $E_T$ jet data in order to determine 
the high  $x$ gluon contribution. In particular, by involving jets in a range of rapidity intervals,  
the $D{\not}O$ measurement allowed to constrain the partons over a much wider $x$ range 
(see fig.~\ref{fig:fig02}: left).  
These new PDFs sets (CTEQ6~\cite{CTEQ6} and MRST01~\cite{MRST01})
represent the most complete information available for Run II jet predictions, 
but are still characterized by a big uncertainty on 
high x gluon distributions (see fig.~\ref{fig:fig02}: right). Run II jet measurements are 
expected to reduce this uncertainty before the LHC era. 
\vspace{-0.4cm}
\section{Preliminary Run II Jet Measurements}
\vspace{-0.2cm}
QCD jet studies at Tevatron are expected to be greatly extended in Run II considering both 
the larger collected data samples and the higher cross section which is related to the increase 
of the center-of-mass energy $\sqrt{s}$ from 1.8 to 1.96 TeV. The improved performances of the 
upgraded CDF and $D{\not}O$ detectors are also expected to play a role. 

\begin{figure}[!htb]
\begin{center}
\includegraphics[width=0.55\linewidth,angle=0]{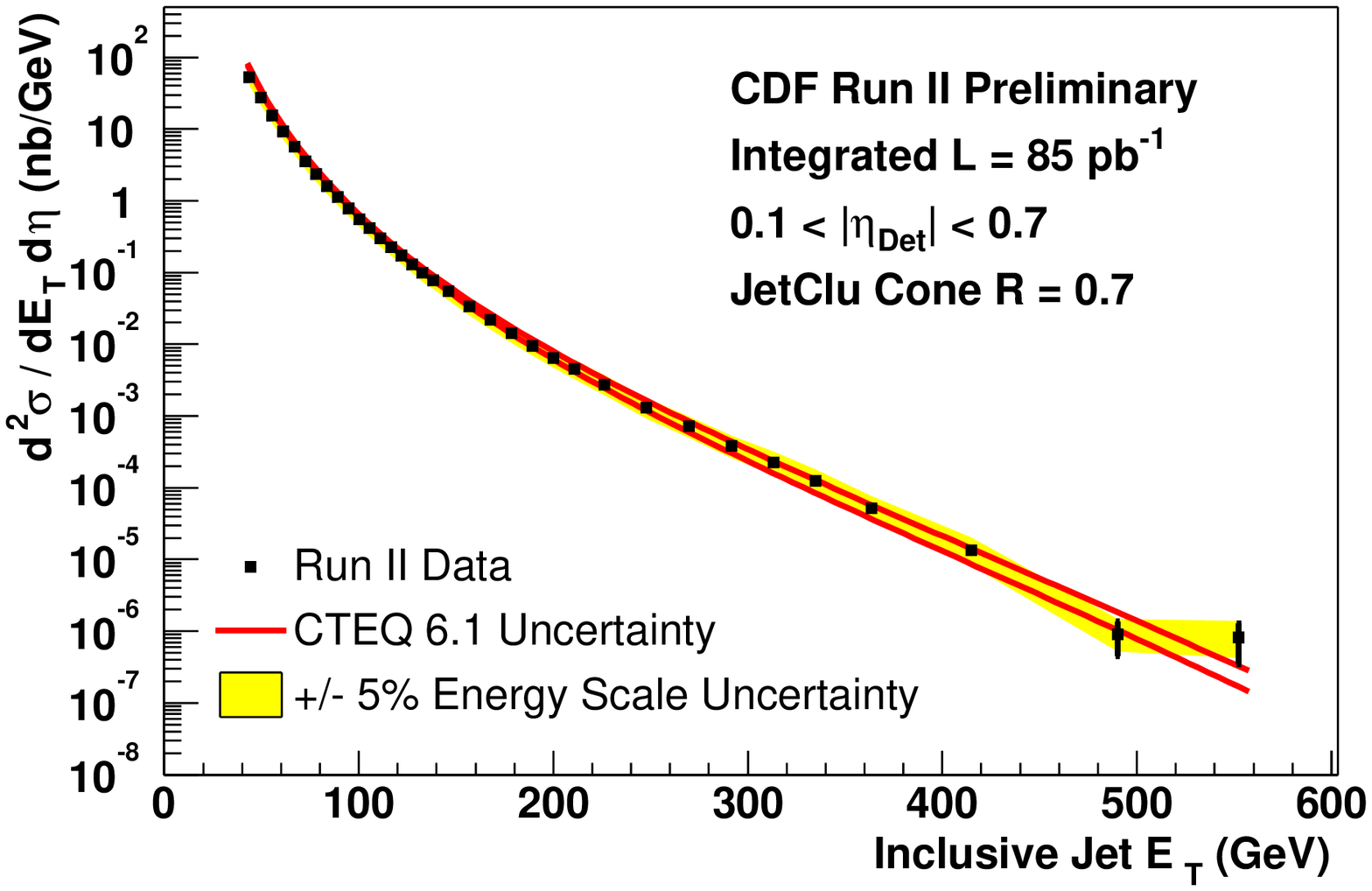}
\includegraphics[width=0.43\linewidth,angle=0]{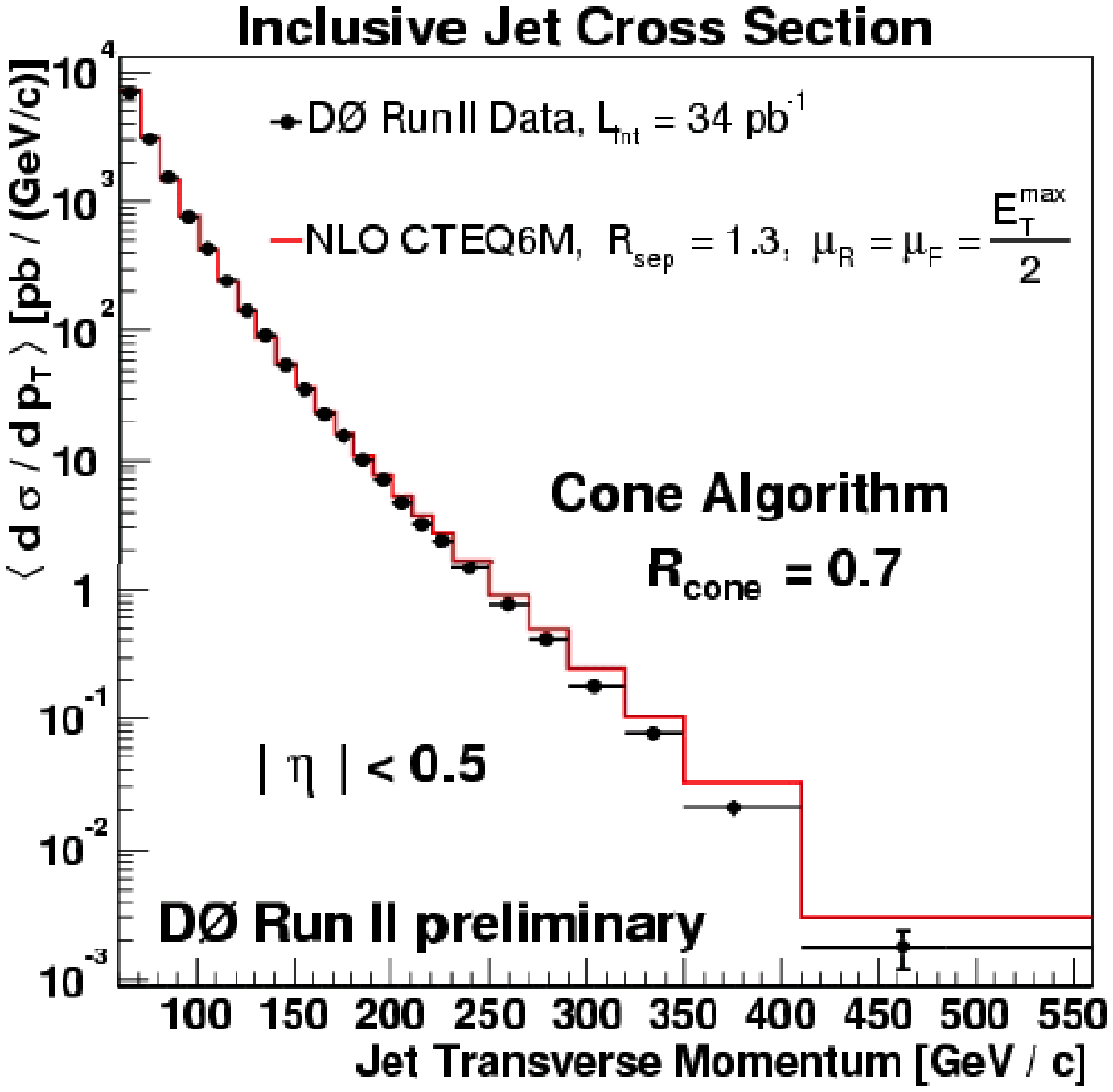}
\vspace{-0.3cm}
\caption {
{\em Inclusive jet cross section as measured in CDF (left)
and $D{\not}O$ (right) respectively analyzing 85 and 34 $pb^{-1}$ of Run II data 
at $\sqrt{s}$ = 1.96 TeV. The results are compared to the NLO pQCD prediction from 
EKS~\cite{EKS} ($\mu$ = $E_{T}^{J}/2$, CTEQ6.1M PDFs) in CDF and JETRAD~\cite{JETRAD} ($\mu$ = 
$E_{T}^{max}/2$, CTEQ6M PDFs) in $D{\not}O$. }
}
\label{fig:fig03}
\vspace{-.5cm}
\end{center}
\end{figure}

Both experiments have recently performed preliminary Run II measurements of the inclusive jet 
and di-jet mass cross sections. 
In order to compare the results at different $\sqrt{s}$, 
CDF adopted the same analysis techniques as in Run I. In particular, the same cone algorithm 
(JetClu~\cite{jetclu}, $R$ = 0.7) and correction procedures (but Run II updated) were used in order 
to reconstruct jets and to properly correct them by taking into account both 
detector and physics effects. The same jet pseudorapidity ranges were also considered.
$D{\not}O$ performed two completely new analyses. For both of them, central ($|\eta|$ $<$ 0.5) jets 
were reconstructed with an ``optimized'' cone algorithm (MidPoint~\cite{midpoint}, $R$ = 0.7) 
and then corrected with a new set of jet corrections as derived from $\gamma$-jet and di-jet 
balancing studies on data~\cite{d0_jet_calib}. 
Results from the inclusive jet cross sections measurements are shown in fig.~\ref{fig:fig03}: 
in both experiments, a reasonably good agreement is found within errors between
data and NLO pQCD theory using the latest CTEQ PDFs sets~\cite{CTEQ6}\footnote{
Similar results were also found by $D{\not}O$ using the MRST01 set~\cite{MRST01}.
}.
A reasonable agreement was also observed, within errors, between data and theory for 
the di-jet mass cross section measurements.
Systematic uncertainties were found to be dominanted by the error on jet energy scale in 
data and on high x gluon distribution function in theoretical calculations. 

Refined analysis techniques, a sensible increase in Run II data samples and a better 
understanding of Run II CDF and $D{\not}O$ detectors performances, are expected to update and 
improve the very preliminary results reported here.

\vspace{-0.4cm}
\begin{thebibliography}{99}
\vspace{-0.2cm}
%
\bibitem{CDF_I}
F. Abe {\em et al.}, (CDF Coll.), Nucl. Instr. and Meth. A {\bf 271}, 
387 (1988).  
\vspace{-0.3cm}
%
\bibitem{D0_I} 
S. Abachi {\em et al.}, ($D{\not}O$ Coll.), Nucl. Instr. and Meth. A
{\bf 338}, 185 (1994).
\vspace{-0.3cm}
%
\bibitem{CDF_old_incl_jet}
F. Abe {\em et al.}, (CDF Coll.), Phys. Rev. Lett. {\bf 77}, 438 (1996).
\vspace{-0.3cm}
%
\bibitem{CTEQ_PDF}
H.L. Lai {\em et al.}, Phys. Rev. D {\bf 55}, 1280 (1997). \\
H.L. Lai {\em et al.}, Eur. Phys. J. {\bf C12}, 375 (2000).
\vspace{-0.3cm}
%
\bibitem{CDF_new_incl_jet}
F. Affolder {\em et al.}, (CDF Coll.), Phys. Rev. D {\bf 64}, 032001 (2001).
\vspace{-0.3cm}
%
\bibitem{D0_new_incl_jet}
B. Abbott {\em et al.}, ($D{\not}O$ Coll.), Phys. Rev. Lett. {\bf 86}, 1707 (2001).
\vspace{-0.3cm}
%
\bibitem{CTEQ6}
J. Pumplin {\em et al.}, JHEP, 0207 (2002). \\
D. Stump {\em et al.}, hep-ph/0303013 (2003).  
\vspace{-0.3cm}
%
\bibitem{MRST01}
A.D. Martin {\em et al.}, hep-ph/0110215 (2001);  hep-ph/0211080 (2002).
\vspace{-0.3cm}
%
\bibitem{jetclu}
F. Abe {\em et al.}, (CDF Coll.), Phys. Rev. D {\bf 45}, 1448 (1992).
\vspace{-0.3cm}
%
\bibitem{midpoint}
G. C. Blazey {\em et al.}, hep-ex/0005012 (2000).
\vspace{-0.3cm}
%
\bibitem{d0_jet_calib}
B. Abbott {\em et al.}, ($D{\not}O$ Coll.), Nucl. Instr. and Meth. A {\bf 424}, 352 (1999).
\vspace{-0.8cm}
%
\bibitem{EKS}
S.D. Ellis {\em et al.}, Phys. Rev. Lett. {\bf 64}, 2121 (1990).
\vspace{-0.3cm}
%
\bibitem{JETRAD}
W.T. Giele {\em et al.}, Phys. Rev. Lett. {\bf 73}, 2019 (1994).
\vspace{-0.3cm}
%
\end{thebibliography}
\end{document}